\documentclass[useAMS,usenatbib,fleqn]{mn2e}

\usepackage{times}

\usepackage{amsmath}
\usepackage{upgreek}
\usepackage{graphicx}

\renewcommand{\vec}[1]{\ensuremath{\bmath{#1}}}   

\title[New 3D Maser Code Applied to Flares]{A New 3D Maser Code Applied to Flaring Events}
\author[M. D. Gray, L. Mason and S. Etoka]
{M. D. Gray$^{1}$, L. Mason$^{1,2}$ and S.Etoka$^{1}$\\
$^{1}$Jodrell Bank Centre for Astrophysics, School of Physics and Astronomy, University of Manchester,
M13 9PL, UK\\
$^{2}$Pendleton Sixth Form Centre, Salford City College, M6 7FR, UK}
\begin{document}

\date{}

\pagerange{\pageref{firstpage}--\pageref{lastpage}} \pubyear{2017}

\maketitle

\label{firstpage}

\begin{abstract}
We set out the theory and discretization scheme for a new finite-element computer code,
written specifically for the simulation of maser sources. The code was used to compute
fractional inversions at each node of a 3-D domain for a range of optical thicknesses.
Saturation behaviour of the nodes with regard to location and optical depth were broadly
as expected. We have demonstrated via formal solutions of the radiative transfer equation
that the apparent size of the model maser cloud decreases as expected with optical depth as viewed
by a distant observer. Simulations of rotation of the cloud allowed the construction of
light-curves for a number of observable quantities. Rotation of the model cloud may be
a reasonable model for quasi-periodic variability, but cannot explain periodic flaring.
\end{abstract}

\begin{keywords}
masers -- radiative transfer -- radio lines: general -- radiation mechanisms: general 
-- techniques: high angular resolution -- ISM: lines and bands.
\end{keywords}

\section{Introduction}
\label{intro}

Astrophysical masers are observed in a variety of source types, particularly
Galactic high-mass star-forming regions, the circumstellar envelopes of highly
evolved stars, and the nuclei of some external galaxies (megamasers). In 
most environments, masers are generally considered to form
in gaseous clouds that contain a population of the active molecular species.
Such clouds are often modelled as spheres, cylinders or slabs, the last possibly
being a better approximation to shock-compressed regions than either spheres
or cylinders. However, in general we expect maser-forming clouds to have
irregular shapes without exploitable symmetry, and the medium as a whole to
consist, perhaps, of an assembly of similar clouds embedded in a surrounding medium
that is usually considered to be significantly more diffuse than the maser clouds
themselves.

The current work introduces a new radiative transfer code, written specifically for masers, 
that can treat clouds of arbitrary geometry. The code uses finite element discretization,
and population solutions are determined at each node of the model. An obviously
interesting problem that can be addressed directly by the new code is that of the
beaming of maser radiation, and the effect that this has on observations, based on
the position of the observer. An extension of this beaming problem, for a single
cloud, is an analysis of the periodic, or near periodic, flaring that may result
from the rotation of an irregular maser cloud through an observer's line of sight.

\subsection{Non-Filamentary Models of Masers}
\label{ss:comput}

General equations for maser radiative transfer, including propagation in more
than one dimension, have a rather long history, extending back at least as far as
the work of \citet{1972ApJ...174..517G}, who developed analytic expressions for
the apparent sizes of spherical and cylindrical sources under a number of specific
saturation conditions. \citet{1973ApJ...180..647L} developed an approximate analytic
solution of the radiative transfer equation for a spherical maser, by using rectangular
lineshapes, and checked the accuracy of the solution against numerical models. The work
of \citet{1973ApJ...182..711L} introduced a new analytical approximation for a
spherical maser, and also introduced radial inhomogeneity, including a radial
velocity field. A different approach to the problem \citep{1976A&A....50..231B},
including partial velocity redistribution (PVR) via different absorption and emission line shape
functions, resulted in an approximate solution for a spherical system in which the
maser gain coefficient had dependence on both radius and angle.

Frequently discussed
parameters in the context of masers with non-filamentary structure are the degree
of line shape re-broadening under saturation and the behaviour of the beam
solid angle as a function of frequency.
These are in part linked to whether
a model assumes complete (CVR) or negligible (NVR) velocity redistribution.
\citet{1992ApJ...393L..37N} discussed possible problems with a `standard' spherical
maser model, as adopted for example by \citet{1985ApJ...290..433A,1990ApJ...363..638E}. Whilst the
standard model assumes CVR, strongly radially beamed radiation
and a gain coefficient that is only a function of radius, Neufeld's model assumed
NVR and arbitrary ray directions. The principal conclusions
from numerical results computed by \citet{1992ApJ...393L..37N} are, firstly, that the beam solid
angle falls somewhat with frequency out to 1 Doppler width under NVR, whereas it rises
dramatically under the standard model, and, secondly, that the NVR model predicts substantial
amplification for impact parameters greater than the saturation radius (dividing the
unsaturated core from the saturated exterior), but the standard model does not.

Results of more modern NVR computational
models \citep{1994ApJ...424..991E,2003ApJ...598..357W} for spheres and thin discs
demonstrate that, under strong saturation, line profiles of masers of these geometries
re-broaden to a near-gaussian form in a manner very similar to a filamentary maser.
The only really significant departure of the results in \citet{1994ApJ...424..991E} 
from the standard model for
spheres is in the behaviour of the apparent maser size as a function of frequency: whilst
the standard model shows a monotonic rise, the more sophisticated model can, under
some conditions of saturation, show a small reduction in size at frequencies below
1 Doppler width, whilst strong increase in apparent size with frequency does not
start until larger shifts from line centre. \citet{2003ApJ...598..357W} links the frequency 
at which rapid size increase begins to an intensity level of order 0.1 times that at
line centre, rather than to any specific frequency.

A spherical maser model of 40 levels of o-H$_2$O \citep{2013A&A...553A..70D} is likely the
most sophisticated of the spherical models. It separates maser and `thermal' lines into
groups that are treated, respectively, under NVR and CVR. Radiative transfer solutions
employ the short-characteristic method by \citet{1987JQSRT..38..325O} and use is also 
made of work by \citet{1993ApJ...407..620A}. \citet{2013A&A...553A..70D} do not 
comment on the frequency-dependence of spot-sizes in their model, but do show that 
their accurate predictions of overall
maser gains may differ by at least an order of magnitude from predictions that used
less accurate methods. Mention should also be made of a spherical shell solution for
the OH 1612-MHz maser in evolved stars \citep{1992MNRAS.258..159S}. 

In recent years, more general radiative transfer codes, for 
example, {\sc lime} \citep{2010A&A...523A..25B}
have been developed that allow pseudo-random point generation by sampling some physical
model, with subsequent Delaunay triangulation of the point distribution in a similar manner
to the current work. However, masers, particularly of strongly saturating intensity, are
not treated accurately by {\sc lime}. {\sc lime} has recently been employed
to study methanol emission in model T~Tauri discs \citep{2016MNRAS.460.2648P}, but no 
significantly saturating maser emission was generated, so possible saturation problems 
were not investigated in that work.

\subsection{Maser Flares}
\label{ss:flares}

Maser sources were discovered, at the time of writing, 52 years ago, and
are known to vary on timescales ranging from 1000\,s \citep{1992ApJ...387L..73S} to decades:
One type of variability
is flaring, examples of which are known from at least the species OH \citep{1979ApJ...228L..79C}, 
H$_2$O \citep{1973ApJS...25..393S,1979IAUC.3415....2A}, CH$_3$OH \citep{2005MNRAS.356..839G}, 
NH$_3$ \citep{1988IAUC.4537....1M} and H$_2$CO \citep{2007ApJ...654L..95A}. The sources of 
these flares include star-forming regions, evolved stars and megamasers.
Variability surveys have demonstrated that maser time-dependence, at least for the
Class~II methanol masers in massive star-forming regions, can have a wide variety of patterns, 
including monotonic increase and decay, and flaring behaviour \citep{2004MNRAS.355..553G}. 
Flaring may be aperiodic, quasi-periodic
or periodic. \citet{2004MNRAS.355..553G} found seven periodic methanol maser sources
with periods beween 132 and 668\,d. The great majority of variability is not periodic, at
least on timescales that have so far been studied. Periodic variability with shorter
periods ($\sim$24\,d) has also been discovered \citep{2017PASJ...69...59S}.

There appears to be no standard definition of a flare that clearly separates such
behaviour from any other type of variability, either with respect to duration,
or maximum-to-minimum flux ratio. However, typical rise times are of order a few
weeks to months, followed by a usually somewhat slower decay, for example
\citet{1984SvAL...10..307L,1996A&A...315..134E}, but faster behaviour in single
spectral features (hours to days) has been observed \citep{1993AAS...182.7507A}. 
Flux ratios between
flaring maximum and a non-flaring state vary from apparently infinite (the original or
non-flaring state is not detectable) to values below 2:1. Water maser flares in Orion
\citep{1982SvAL....8...86S} have yielded the highest brightness temperatures (up to
8$\times$10$^{17}$\,K) ever recorded for astrophysical masers. Other commonly observed
properties include strong polarization and narrowing (broadening) of spectral lines
as the flare brightens (fades).

Many theories have been put forward to explain flares, and these are often
specific to a source or outburst. Some possibilities include binary orbits 
affecting infra-red pumping \citep{2003MNRAS.339L..33G}, more general variations
in radiative pumping \citep{1996A&A...315..134E}, shock compression and heating
of a cloud \citep{1982SvAL....8...86S}, superradiant (non-maser)
bursts \citep{2017SciA....3E1858R}, and the passage of a rotating foreground cloud
through the line of sight to another maser cloud \citep{1998ApJ...509..256B}.
More recent developments have included detection of correlated variability in different
maser transitions, and even between different species, for example periodic
and alternating flares in water and
methanol \citep{2016MNRAS.459L..56S}, said to be due to cyclic accretion
instabilities in a protostellar disc.

We introduce below, a first essay at a model that can address flaring resulting
from at least the following scenarios: cloud rotation, superimposition of two
or more clouds in the line of sight, changes in the maser optical depth or
pumping strength, and variability in the amplified background radiation.

\section{Theory}
\label{s:theo}

The coupled equations of radiative transfer and saturation of the inversion have been
solved, in this work, in a geometrically irregular domain with finite-element
discretization. In any solution, the fractional inversion, that is the ratio
of the saturated to the unsaturated inversion, of a two-level model was
determined at every node of the domain. The radiation observable from the domain at any
distant viewpoint was calculated by amplifying background radiation through a 
domain solution along approximately parallel ray paths. Flaring behaviour resulting
from rotation of the domain was simulated by choosing a set of viewpoints that form
a circle around the domain.

\subsection{Domain Generation}
\label{ss:domain}

Domains were generated by Delaunay triangulation of the space formed by a weighted
random distribution of nodal points. Where random numbers have been used in this work,
the generating routine is based upon the `{\sc ran2 }' subroutine from \citet{1992nrfa.book.....P}
that generates uniform deviates with a repeat period longer than $2\times10^{18}$.
The initial set of points were sorted into a
boundary set and an internal set on the basis of their distance from the origin of the
model. In the models used in this work, the boundary set was formed from the
more distant 10 per cent of points.
The Delaunay triangulation itself follows a description in \citet{zienkiebook}: A bounding
cuboid was initially divided into six tetrahedral elements. In this work, all elements
are of the linear simplex type, so in 3-D they are tetrahedra with four nodes placed
at the vertices. An initial discretization was then made between the set of boundary points 
and the eight corners of the bounding cuboid to form the `box mesh'. At the end of this
process, the cuboid support was removed by deleting all elements that included, as nodes, at
least one corner of the bounding cuboid.

At the next stage of triangulation, the remaining nodes, that is those not in the
boundary set, were inserted into the irregular shape, the domain, formed by the boundary nodes,
with a resulting increase in the overall number of elements at each insertion. If a point
from the original weighted-random list fell outside the domain, it was ignored. 
Additional machine-generated points could be inserted in situations where the original
elements had poor aspect ratios. The
result of the triangulation was the usual pair of triangulation tables: the node
definition table that records the cartesian 3-D coordinates of each node, and the
element definition table that records the global numbers of the member nodes of each
element, and the neighbour elements on each face. A neighbour number of zero is
stored for external faces. The cartesian coordinates of all nodes were guaranteed to
be in the range $-1..1$ along all three axes.

A table of edges (internal and external) of the final domain
is not required for the computational solution, but was stored for graphical
purposes only. A geometrical domain, constructed as above, may be modified by introducing 
a set of physical conditions at each node.

With no loss of generality, a table of shape-function coefficients can be calculated,
since these depend only upon coordinates of the nodes of each tetrahedral element. Since
there is one shape-function for each node, and each function has four coefficients,
a total of 16 coefficients had to be stored per element. The method of calculation was
standard, with the coefficients being formed from determinants of nodal positions,
drawn from the nodal definition table, see for example \citet{zienkiebook}. The shape
function for local node $j$ (in the range 1..4) in terms of the coefficients computed above, 
takes the form,
\begin{equation}
f_j(x,y,z) = a_j + b_j x + c_j y + d_j z,
\label{eq:defshape}
\end{equation}
so that any scalar field $\phi(x,y,z)$ can be represented within an element as
\begin{equation}
\phi (x,y,z) = \sum_{j=1}^4 (a_j + b_j x + c_j y + d_j z) \phi_j = \sum_{j=1}^4 f_j(x,y,z) \phi_j,
\label{eq:defscal}
\end{equation}
where the $\phi_j$ are the known nodal values of the field.

Once
computed, the shape-function coefficients were subjected to a sum test: in any
element the sum of the first coefficients, $\Sigma_i a_i = 1$, where $i$ is the
local node index. For the other coefficients $\Sigma_i \xi_i = 0$, where $\xi$ can
be any of the later coefficients, $b,c$ or $d$.

\subsection{Formulation}
\label{ss:form}

Although the geometrical model is sophisticated, the fractional inversion at some
position, $\vec{r}$, within the domain is 
simply the two-level result,
\begin{equation}
\Delta (\vec{r}) = \frac{1}{1 + \bar{j}(\vec{r})},
\label{eq:inversion}
\end{equation}
where $\bar{j} (\vec{r})$ is the mean intensity (angle and
frequency averaged) at position $\vec{r}$ as a multiple of the
saturation intensity, that is the intensity necessary to enforce $\Delta (\vec{r})=0.5$. The
saturation intensity is assumed the same at all positions, which implies that the domain
has uniform physical conditions in the absence of maser radiation. Although we take
eq.(\ref{eq:inversion}) to be valid at an arbitrary position within the domain,
it will usually be computed at one of the nodes, $k=1..K$, of the domain, but this will
not yet be enforced.

If spontaneous emission is ignored, the radiative transfer equation that must be solved
in conjunction with eq.(\ref{eq:inversion}) is
\begin{equation}
di_\nu / ds = \gamma_\nu (s) i_\nu ,
\label{eq:rteq}
\end{equation}
where $i_\nu$ is the specific intensity at frequency $\nu$ as a multiple of the
saturation intensity, and $\gamma_\nu$ is the gain coefficient at the same frequency
that depends upon the position $s$, along the ray
path, only via the inversion and the line shape function,
$\phi_\nu$. In fact, in this model, where pumping conditions are identical for all
nodes, we may write
\begin{equation}
\gamma_\nu (s) = \gamma_0 \Delta (s) \phi_\nu (s),
\label{eq:gcoef}
\end{equation}
where $\gamma_0$ is a constant of the model. The trio of equations,
eq.(\ref{eq:inversion})-eq.(\ref{eq:gcoef}) can be combined into
a single radiative transfer equation, with inclusion of saturation:
\begin{equation}
\frac{di_\nu}{ds} = \frac{\gamma_0 i_\nu(s)\phi_\nu(s)}{1+\bar{j}(s)},
\label{eq:rtcomb}
\end{equation}
where the equation of the straight-line corresponding to any ray can be used to convert
between the $s$ and $\vec{r}$ representations of position.

We assume that the line shape function of the molecular response is always Gaussian
under conditions of negligible saturation. In this limit, the stimulated emission rate
of the maser is vastly smaller than the rate of redistributive processes, such as
kinetic collisions, that shift the response of individual molecules across the line
shape. Two common approximations used in maser studies are the limit of complete 
redistribution, or CVR,
in which a Gaussian response is maintained at any level of saturation, and the limit of
negligible redistribution, or NVR, in which all velocity subgroups behave independently.
We note that CVR is likely a better approximation for weak and modest levels of saturation,
but must eventually become poor under very strong saturation, when stimulated emission
becomes the fastest microscopic process. The present model considers CVR, so that the
molecular response throughout follows the Gaussian form,
\begin{equation}
\phi_\nu (s) = \frac{1}{\pi^{1/2}\bar{W}}\frac{\bar{W}}{W(s)} \exp \left\{
                -\frac{\bar{W}^2 (\tilde{\nu} - \hat{\vec{n}}\cdot \vec{u}(s))^2}{W^2(s)}
                                        \right\},
\label{eq:gaussian}
\end{equation}
where $W(s)$ is the frequency width of the distribution at point $s$ along a ray, $\bar{W}$ is
the mean frequency width over all nodes of the model, $\hat{\vec{n}}$ is a unit vector in the
direction of the ray, and $\vec{u}(s) = \nu_c \vec{v}(s)/(c \bar{W})$ is the dimensionless velocity
of the cloud gas at $s$. The laboratory line centre frequency, $\nu_c$, and vacuum speed of light, 
$c$, have been used to complete this definition. The frequency has also been expressed in
dimensionless form in eq.(\ref{eq:gaussian}), where
\begin{equation}
\tilde{\nu} = (\nu - \nu_c) / \bar{W}.
\label{eq:nudimless}
\end{equation}

The factor of $1/(\pi^{1/2}\bar{W})$ from eq.(\ref{eq:gaussian}) can now conveniently
be combined with $\gamma_0$ in eq.(\ref{eq:rtcomb}) to form a constant
of the model that can be absorbed into the path element
to form the frequency-independent optical depth, $d\tau = [\gamma_0 /(\pi^{1/2} \bar{W})]ds$.
In terms of this new independent variable (along a single ray), the transfer equation,
eq.(\ref{eq:rtcomb}) becomes,
\begin{equation}
\frac{di_{\tilde{\nu}}}{d\tau}
=
\frac{\bar{W} i_{\tilde{\nu}}(\tau)}
     {W(\tau)(1 + \bar{j}(\tau))} \exp \left\{
                -\frac{\bar{W}^2 (\tilde{\nu} - \hat{\vec{n}} \cdot \vec{u}(\tau))^2}{W^2(\tau)}
                                       \right\}.
\label{eq:rtco2}
\end{equation}

In the integral-equation method, used in this work, the aim is to eliminate all radiation
integrals, particularly $\bar{j}(\tau)$, the mean intensity. To this end, we therefore use
eq.(\ref{eq:inversion}) in eq.(\ref{eq:rtco2}) to introduce the inversion, leaving,
\begin{equation}
di_{\tilde{\nu}}/d\tau
=
i_{\tilde{\nu}}(\tau) \Delta(\tau) \eta(\tau) e^{-(\tilde{\nu} - \hat{\vec{n}} \cdot \vec{u}(\tau))^2 \eta^2(\tau)},
\label{eq:rtco3}
\end{equation}
where the new parameter $\eta (\tau) = \bar{W}/W(\tau)$.

Equation~\ref{eq:rtco3} may be integrated along a ray from a point on the boundary,
where the optical depth is assumed to be $\tau_0$, to some arbitrary depth along the ray.
Formally,
\begin{equation}
i_{\tilde{\nu}} (\tau) = i_{BG} \exp \left\{
                 \int_{\tau_0}^{\tau} d\tau' \Delta(\tau') \eta(\tau') 
                   e^{-(\tilde{\nu} - \hat{\vec{n}} \cdot \vec{u}(\tau'))^2 \eta^2(\tau')} 
                         \right\}.
\label{eq:inu}
\end{equation}

The remaining problem, of fundamental importance in a 3-D model, is the evaluation
of the mean intensity at an arbitrary position, so that eq.(\ref{eq:inversion}) may again
be applied to eliminate the radiation integral in favour of the inversion. We first
generate a frequency-averaged intensity by multiplying eq.(\ref{eq:inu}) by the
line shape function and integrating over all possible values of $\tilde{\nu}$. Formally,
\begin{equation}
\bar{i}(\tau) = \frac{1}{\pi^{1/2}} \int_{-\infty}^\infty \phi_{\tilde{\nu}}(\tau) i_{\tilde{\nu}}(\tau) d\tilde{\nu},
\label{eq:ibar}
\end{equation}
noting that the line shape function in the dimensionless variables,
\begin{equation}
\phi_{\tilde{\nu}}(\tau) = \eta(\tau) e^{-\eta^2 (\tau) (\tilde{\nu} - \hat{\vec{n}} \cdot \vec{u}(\tau))^2} ,
\label{eq:dimlessg}
\end{equation}
has the normalization $\int_{-\infty}^\infty \phi_{\tilde{\nu}}(\tau) d\tilde{\nu}=\pi^{1/2}$. The mean 
intensity (frequency and angle-averaged) is obtained
by taking the mean of eq.(\ref{eq:ibar}) over solid angle, with the result
\begin{align}
\bar{j}(\vec{r})
= &
\frac{i_{BG}}{4\pi^{3/2}} \oint d\Omega \eta(\tau) \int_{-\infty}^\infty d\tilde{\nu} 
                      e^{-\eta^2(\tau)(\tilde{\nu} - \hat{\vec{n}} \cdot \vec{u}(\tau))^2} \nonumber \\
             \times &   \exp \left\{
                    \int_{\tau_0}^{\tau} d\tau' \Delta (\tau') \eta (\tau')
                      e^{-\eta^2(\tau')(\tilde{\nu} - \hat{\vec{n}} \cdot \vec{u}(\tau'))^2}
                \right\},
\label{eq:jbar}
\end{align}
where $\vec{r}$ is now the position where many rays, all with different
values of $\tau$ and $\hat{\vec{n}}$, meet, having been traced from points
on the boundary of the domain. Elimination of the mean intensity via eq.(\ref{eq:inversion})
produces the following non-linear integral equation in the inversions at various
positions:
\begin{align}
\Delta (\vec{r})& = \left[ 1 + \frac{i_{BG}}{4\pi^{3/2}}\oint d\Omega \eta(\vec{r})\int_{-\infty}^\infty d\tilde{\nu}
                     e^{-\eta^2(\vec{r})(\tilde{\nu} - \hat{\vec{n}} \cdot \vec{u}(\vec{r}))^2}
                   \right. \nonumber \\
                 &\times \left.
                     \exp \left\{
                            \int_{\vec{r}_0}^{\vec{r}}
                            d\vec{r}' \Delta(\vec{r}') \eta (\vec{r}')
                            e^{-\eta^2(\vec{r}')(\tilde{\nu} - \hat{\vec{n}} \cdot \vec{u}(\vec{r}'))^2}
                          \right\}
                   \right]^{-1}.
\label{eq:nlint}
\end{align}
Note that in eq.(\ref{eq:nlint}), the observation has been made that, whatever the optical 
depth along a particular line of sight, all quantities outside the exponential with braces are
calculated at the position $\vec{r}$. All quantities that are functions
of ray optical depth have been converted
to functions of position, assuming that the equations of all rays (straight lines) are known.
It must therefore be remembered that the set of positions is unique to a ray with
direction along $\hat{\vec{n}}$, each starting from a different point,
$\vec{r}_0 (\hat{\vec{n}})$, but finishing at the same point, $\vec{r}$.

Equation~\ref{eq:nlint} implies that the inversion at $\vec{r}$ depends on the inversions
at all other positions, $\vec{r}'$, within the domain, so the equation is not particularly
useful in this form. However, it does have the significant advantage, pointed out in
\citet{2006MNRAS.365..779E}, that the unknown inversions are always numbers 
between 0 and 1, however large the
maser intensity along any ray may become. In the section that follows, eq.(\ref{eq:nlint}) will be
reduced to a significantly more tractable form for numerical solution.

\section{The Model}
\label{s:model}

The equations from Section~\ref{ss:form} were solved via a non-linear integral equation
method that is the 3-D analogue of the 1-D Schwarzschild-Milne equation from the
theory of stellar atmospheres, see for example \citet{1964ApJ...139..397K}. A computational
slab-geometry implementation of this method has been implemented by \citet{2006MNRAS.365..779E},
and the theory outlined for a more general geometry with finite-element discretization in
\citet{mybook}. 

\subsection{Analytic Frequency Integration}
\label{ss:freqint}

It is possibly advantagous to disentangle the frequency and solid angle integrals
in eq.(\ref{eq:nlint}), so that the final integral equation can be written in terms
of spatial integrals only. This has been achieved here, on an experimental basis, by
completing the frequency integral analytically. In summary, the process involves making
a power-series expansion of the main exponential (the one written with braces in
eq.(\ref{eq:nlint})), developing an expression for the term in an arbitrary power, and
separating this expression into parts dependent on, and independent of, frequency.
The frequency integral can then be carried out term by term via a standard
formula, no. 7.4.32 of \citet{absteg}. Details are deferred to Appendix~\ref{app:freq},
but if the cloud is assumed to be uniform, the frequency-integrated expression for
the $n$th term in the expansion has the relatively simple form,
\begin{equation}
q_n
=
\frac{\pi^{1/2}}{n!(n+1)^{1/2}} \left[
                               \int_{\vec{r}_0}^{\vec{r}} \Delta(\vec{r}') d\vec{r}'
                             \right]^n.
\label{eq:termn}
\end{equation}
The modified form of eq.(\ref{eq:nlint}) contains an infinite sum of terms of the
form given in eq.(\ref{eq:termn}), but now contains only the spatial integrals
over solid angle (effectively a sum of rays) and the integral over positions along an
individual ray that appears in eq.(\ref{eq:termn}):
\begin{equation}
\Delta (\vec{r}) \! = \!\! \left[ \! 1 + \frac{i_{BG}}{4\pi}\!\oint \!d\Omega \!\sum_{n=0}^\infty
                            \frac{\pi^{1/2}}{n!(n+1)^{1/2}} \! \left(
                               \int_{\vec{r}_0}^{\vec{r}} \!\! \Delta(\vec{r}') d\vec{r}'
                                                        \right)^n \!
                   \right]^{-1} .
\label{eq:nluni}
\end{equation}
The reduction of eq.(\ref{eq:nluni}) to a finite number of nodal positions is
considered below; a uniform cloud will from now on be taken as standard.

\subsection{Discretization}
\label{ss:discrete}

The aim here is to put eq.(\ref{eq:nluni}) into a form in which the fractional inversion is evaluated
only at nodal points of the domain. The nodal points are assumed to have already been
distributed amongst a set of simplex finite elements (see Section~\ref{ss:domain} above).
First, the inner integral in eq.(\ref{eq:nluni}) that represents a ray path is broken into
sections, each of which lies entirely within one finite element:
\begin{equation}
\int_{\vec{r}_0}^{\vec{r}} \Delta(\vec{r}') d\vec{r}'
= \sum_{\{ j \}} \int_{\vec{r}_{0j}}^{\vec{r}_{1j}} \Delta(\vec{r}') d\vec{r}',
\label{eq:rayel}
\end{equation}
where a ray consists of a set $\{ j \}$ of elements along the path, each
with an entry point $\vec{r}_{0j}$ and an exit point $\vec{r}_{1j}$.

The variation of any physical quantity, including the unknown inversion, within
an element may be represented by the element shape functions in terms of its
nodal values. Mathematically, if an element has $P$ shape functions and $P$ nodes, then at
any point $\vec{r}'$ within element $j$, the inversion is, from eq.(\ref{eq:defscal}),
\begin{equation}
\Delta(\vec{r}') = \sum_{p=1}^P f_p(\vec{r}') \Delta_p^j ,
\label{eq:shape}
\end{equation}
where $f_p$ is the shape function corresponding to node $p$, and $\Delta_p^j$ is
the inversion at node $p$ of element $j$.
Combining equations eq.(\ref{eq:rayel}) and eq.(\ref{eq:shape}) completes the
discretization of the ray integral, since the element shape functions are
known:
\begin{equation}
\int_{\vec{r}_0}^{\vec{r}} \Delta(\vec{r}') d\vec{r}'
= \sum_{\{ j \}} \sum_{p=1}^P \Delta_p^j \int_{\vec{r}_{0j}}^{\vec{r}_{1j}} f_p(\vec{r}') d\vec{r}'
= \sum_{\{ j \}} \sum_{p=1}^P \Delta_p^j F_p^j
\label{eq:raydisc}
\end{equation}
and the integral can be carried out to yield the final version on the
right-hand side of eq.(\ref{eq:raydisc}).

As inversions in eq.(\ref{eq:nluni}) can be specified at an arbitrary position, we choose
an arbitray node with global index number, $i$, reducing this equation to
\begin{equation}
\Delta_i \! = \!\! \left[ \! 1 \! + \! \frac{i_{BG}}{4\pi}\!\oint \!d\Omega \!\sum_{n=0}^\infty
                            \frac{1}{n!(n+1)^{1/2}} \! \left(
                                \sum_{\{ j \}} \sum_{p=1}^P \Delta_p^j F_p^j
                                                       \! \right)^{\!\!\!n} 
                   \right]^{-1},
\label{eq:fivethree}
\end{equation}
noting that only the solid-angle integral remains in a continuous form. This
integral must be replaced by a finite sum over the rays converging on node $i$
from the boundary. In fact, a double sum is used: one sum is over boundary
faces (index $q$, total $Q$), and the other, over the number of rays assigned
to start from each of these faces (index $b$ with $B$ rays per boundary face).
An algorithm must be supplied (see Section~\ref{sss:algomain} below) that specifies an
area $a_{bq}$ per ray. The solid angle element associated with each ray is
then
\begin{equation}
\delta \Omega_{bq} = a_{bq} / s_{bq}^2 = a_{bq} / (\vec{r} - \vec{r}_{0bq})^2 ,
\label{eq:solang}
\end{equation}
where $\vec{r}_{0bq}$ is the position where a specific ray enters the domain.
An initial discretized form of eq.(\ref{eq:fivethree}) is therefore
\begin{equation}
\Delta_i \! = \!\! \left[ \! 1 \!+\! \frac{i_{BG}}{4\pi}\! \sum_{q=1}^Q \sum_{b=1}^B \!\frac{a_{bq}}{s_{bq}^2} \!\!\sum_{n=0}^\infty
                            \!\frac{1}{n!(n\!+\!1)^{1/2}} \!\! \left(
                                \!\sum_{\{ j \}} \sum_{p=1}^P \Delta_p^j F_p^j \!\!
                                                        \right)^{\!\!\!n} 
                   \right]^{\!-1}.
\label{eq:rawdisc}
\end{equation}

The significant problem that remains with eq.(\ref{eq:rawdisc}) is that node $i$ is specified
globally, whilst the other nodes are specified locally, with two indices, corresponding
to an element number and node number of 1 to 4 within that element.
A single global node may be a member of more than one element. The aim is to
obtain a single coefficient that represents the contribution of the saturated
inversion of a single global node, via some set of rays, to the inversion at
the target node, $i$.

A first step is to re-write the double-sum over the path (the set $\{ j \}$ of
elements and $p=1..P$ nodes per element) as a single sum over nodes in terms of their
global numbers. Re-define $j=1..M_{bq}$ as the index of unique nodes that appear (possibly more
than once each) in the set of elements along an individual ray path.
Each of these nodes now contributes a saturating effect via a coefficient $\Phi_{G(j)}$ that is
some combination of the relevant $F_p^j$, and $G(j)$ is the global node number
of a node $j$ from the list $j=1..M_{bq}$. A similar procedure can be applied to the double sum
over external faces and rays per face, but it is easier, since each ray is unique. The
result is,
\begin{equation}
\Delta_i  - \! \left[ \! 1 \!+\! \frac{i_{BG}}{4\pi} \!\!\sum_{q'=1}^{Q'} \!\frac{a_{q'}}{s_{q'}^2} \!\sum_{n=0}^\infty
                            \!\frac{1}{n!(n\!+\!1)^{1/2}} \! \left(
                                \sum_{j=1}^J \Phi_{j,q'} \Delta_j \!
                                                        \right)^{\!\!\!n} \!
                   \right]^{-1} \!\!= 0,
\label{eq:globdisc}
\end{equation}
where $Q'$ and $J$ are now the respective total numbers of rays and nodes in the model. The adoption
of $J$ amounts to extending the single-ray limit of $j=M_{bq}$ (see above) to include all nodes in
the model, since a single node may contribute to more than one element on a ray to a given target 
node, and to more than one ray. Many ray and node pairs contribute zero saturating effect at 
target node $i$, and only non-zero contributions are stored.
The new coefficient, $\Phi_{j,q'}$, now expresses this saturating effect of ray $q'$ and
global node $j$ on the target node $i$, noting that it will in general be the sum of several 
of the original $F$-coefficients from eq.(\ref{eq:rawdisc}). Since the algorithm for 
solving eq.(\ref{eq:globdisc}) is iterative, and the iteration sequence is based on its 
residual, a current estimate for the set of $\Delta_j$ is always available.

It is possible to
use the multinomial theorem to re-write the power-of-$n$ bracket
in a form in which the individual element sums (over index $p$) are written as
a finite product. The second step is to apply the multinomial theorem again to each
of these element sums, each of which has $P$ terms, where typically $P=4$. However, we
do not consider this anlytical scheme further.

\subsection{Algorithms}
\label{ss:algo}

\subsubsection{Main code}
\label{sss:algomain}

The main routine, responsible for solving the set of non-linear algebraic
equations represented by eq.(\ref{eq:globdisc}), uses a Levenberg-Marquardt (LM) two-stage
algorithm as implemented by \citet{Amini2015341}. This routine was successful unsupported for modest
levels of saturation, but its performance was much improved under more
challenging conditions by the use of a simple least-squares convergence
accelerator \citep{1974JChPh..61.2680N}, usually applied every four LM iterations.

As the model introduced in Section~\ref{s:theo} is simply scalable in optical depth,
solution generally proceeded by solving eq.(\ref{eq:globdisc}) for the same geometrical
domain many times, progressively increasing the optical depth and therefore the degree of
saturation. For optical depths lower than $\sim$ 3, it was found to be effective to
simply use the nodal populations from the previous solution as the starting
estimate for the new, thicker, model. At larger optical depths, a small number of 
previous solutions, typically 3-5, were used to extrapolate to a new starting estimate. 
The extrapolation algorithm was of polynomial type, and based on the `{\sc polint}' 
routine by \citet{1992nrfa.book.....P}.

For the solution of eq.(\ref{eq:globdisc}), the starting positions of rays were based
on a `celestial sphere' structure, centered on the domain origin, but with a significantly
larger radius, typically $r=5$, compared to the $r<1$ of the domain. To allocate rays
with approximately equal solid angle, this external sphere was first partitioned into
an icosahedral figure, such that the vertices of the icosahedron were coincident with
the celestial sphere. The triangular faces were progressively subdivided into smaller
triangles of equal area, with one ray starting at each vertex. It should be noted that a final
back-projection of the intial ray positions onto the celestial sphere involves a small distortion
from an ideal equal-area distribution of rays on the sphere.

For formal solutions, rays forming a narrow cone, pointing towards
a distant observer, were traced through the domain with
known nodal populations. The base of this cone was a disc of diameter greater than
the length of the maximum chord through the domain. The details of ray tracing through
the domain are discussed in Section~\ref{sss:raystuff} below, but the origin positions
of the rays on the source disc were computed, with the goal of an equal-area distribution,
according to a partition algorithm by \citet{Beckers2012275}.

\subsubsection{Ray Tracing}
\label{sss:raystuff}

Since the ray intensities have been entirely eliminated from eq.(\ref{eq:globdisc}), they
do not need to be explicitly calculated, but the route of a ray through the domain matters
fundamentally through the coefficients $\Phi_{j,q'}$. For each target node, all entry
and exit points of the domain are determined. To do this, a subset of boundary elements
is identified: those that have at least one face with a zero (external) neighbour value.
For each external face, an intersection point of the extended plane of the face with the 
ray is computed, and this point is checked to see if it is within the external face, 
in which case a valid intersection is recorded.
Such intersections must occur in pairs, defining one entry, and one exit, point for
a ray segment within the domain. All segments that have an entry point more distant than
the target node from the ray source are discarded. The remaining set of segments must
include a final segment that has an external entry face, but no exit, since the ray
terminates on the target node.

For each segment, the set of member elements is computed: unless the ray reaches the
target node, it must leave the current element via a different face to the one by
which it entered. The exit face must have an outward normal vector that points into the
same hemisphere as the unit vector of the ray, and must have an intersection point of 
its extended plane with the ray that is within the face. The exit face is chosen as the 
closest of those that satisfy the above criteria. The new element into which the ray 
passes, and the entry face of the new element are known immediately from the neighbour 
information. The above process is therefore repeated until element and node membership 
of the ray segment is completed either by exit through a face with
neighbour value 0, or contact with the target node. Element sets for complete ray paths
can then be constructed by combining all segments between the first entry to the domain
and the target node.

The integral in eq.(\ref{eq:raydisc}) that forms the coefficient $F_p^j$ can be
carried out analytically for each element along the ray path with the result,
\begin{align}
F_p^j & = s(a_p + b_p x_{0j} + c_p y_{0j} + d_p z_{0j})\nonumber \\
     & + (s/2) (b_p x_j + c_p y_j + d_p z_j) ,
\label{eq:sfnt}
\end{align}
where the $a_p,b_p,c_p,d_p$ are the shape-function coefficients for (local) node $p$,
as defined in eq.(\ref{eq:defshape}).
The vector $\vec{r}_j = \vec{r}_{1j} - \vec{r}_{0j}$ is the ray path through element $j$
with components $(x_j,y_j,z_j)$ from the entry point $\vec{r}_{0j}=(x_{0j},y_{0j},z_{0j})$
to exit point $\vec{r}_{1j}$. The scalar distance is $s = |\vec{r}|$. By keeping track of the global
node numbers, a sum of the $F_p^j$ can be accumulated for each global node along the
ray, and on completion of the ray to its target, this sum becomes the $\Phi$ coefficient
from eq.(\ref{eq:globdisc}). It should be noted that these coefficients need to be
computed only once per model, comprising a set of rays and a geometrical domain, and
can simply be drawn from memory during the LM iterations.

\section{Results of Computations}
\label{s:comput}

The results discussed here were computed for a domain of modest size, but still
of sufficient scale to provide a useful test of the code. An initial set of
250 nodes was generated with a spherical distribution and even weighting by
volume. After Delaunay triangulation, the final domain consisted of 202 nodes,
of which 27 formed the boundary subset, generating 1177 linear tetrahedral
elements. As expected for 4 nodes per element, the total number of shape-function
coefficients was 4708. The largest estimated fractional error in any computed shape-function
coefficient was $1.688\times 10^{-13}$.

\begin{figure}
  \includegraphics[bb=55 450 595 795, width=94mm,angle=0]{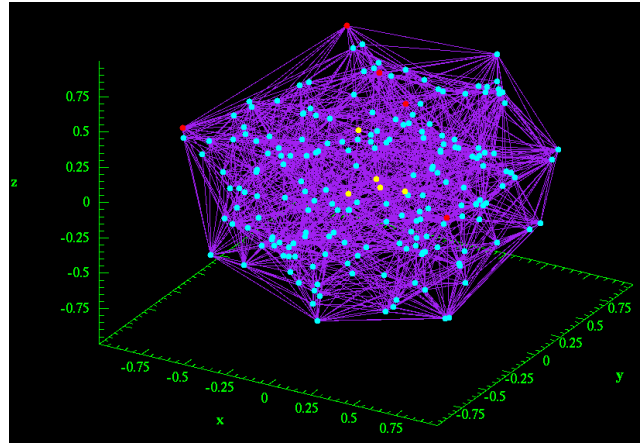}
  \caption{The geometrical domain of 1177 elements, formed from 202 nodes. The nodes
are represented as cyan dots,
except for the five that become the most saturated set (red) and the five that
saturate least (yellow).}
\label{f:domain1}
\end{figure}

The geometrical domain defined above was traversed by a total of 1442 rays towards
each target node, so that a total of 291284 rays were used. The ray-tracing algorithms
then found a total of 3777276 non-zero saturation coefficients (of the $\Phi$-type,
see Section~\ref{sss:raystuff}). A background radiation specific intensity of
$i_{BG}=1.0\times 10^{-5}$, where the saturation intensity is 1.0, was used for
all rays. A view of the geometrical domain is shown in Fig.~\ref{f:domain1}, noting
that the random selection of points, and selection of the boundary set,
produced a significantly non-spherical cloud from an initially spherical point distribution.

\subsection{Nodal Solutions}
\label{ss:nodalsol}

Nodal solutions for the inversion were obtained by solving eq.(\ref{eq:globdisc})
for the domain displayed in Fig.~\ref{f:domain1} for progressively increasing  values of the
model optical depth. Note that since the scalar distance $s$ is a common multiplier
of all parts of $F_p^j$ in eq.(\ref{eq:sfnt}), multiplication of these, or the node-summed
$\Phi$ forms, by an extra scaling factor was sufficient to change the effective optical
depth. After the first solution, where the initial guess at the fractional inversions was 1.0
for all nodes (unsaturated), initial guesses were based upon one or more previous
solutions. Nodal solutions were eventually computed for optical depth multipliers
between 0.5 and 13.5. These multipliers correspond to actual maximum maser depths that 
are approximately twice as large, as the dimensionless outer radius, rather than the 
diameter, of the original point distribution was 1.0. All successful solutions had a 
largest inversion residual smaller than 10$^{-8}$. Convergence problems eventually made 
progress to higher optical depth multipliers
prohibitively slow beyond the upper limit of 13.5. However, it can be argued that
the CVR approximation is already rather poor with the corresponding levels of 
saturation, and a switch to an approximation with weaker redistribution should be made.

\begin{figure}
  \includegraphics[bb=145 60 410 230, scale=1.37,angle=0]{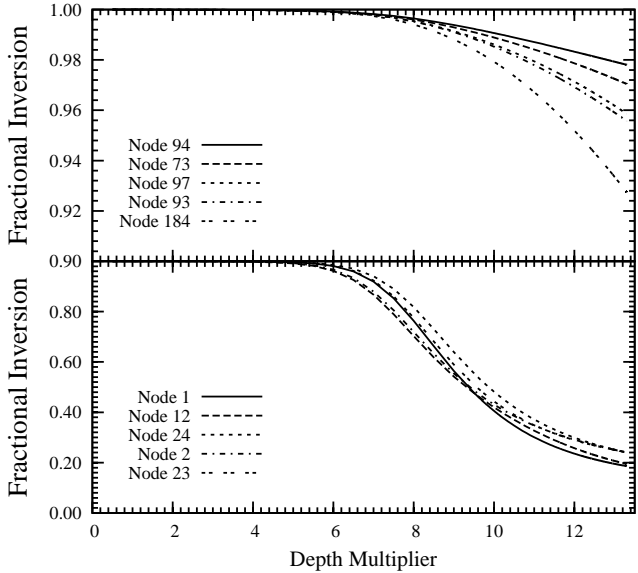}
  \caption{The populations in the five most saturated nodes (lower panel) and the
five least saturated nodes (upper panel) as a function of the optical depth multiplier.}
\label{f:popsat}
\end{figure}

To demonstrate the effects of increasing saturation (with optical depth) we
show in Fig.~\ref{f:popsat} the inversions in the five most saturated, and
the five least saturated, nodes
as a function of the optical depth multiplier. The definitions of most and
least saturated refer to the model at the highest optical depth achieved, as
some swapping of the ordering with saturation does occur: for example, it is easy to
see that Node~1, the least inverted node at maximum depth, does not become so
until the depth multiplier reaches $\sim$8.5. Broadly speaking, the nodal inversions
in Fig.~\ref{f:popsat} behave as expected. For the most saturated nodes, there is
a rapid, almost linear, fall in the inversion with depth between depths of 7 and
10, followed by a distinct slowing of the saturation as the inversions fall
below about 0.3. For the least-saturated set, all inversions remain above 0.9
over the range of depths investigated, and the phase of rapid population drop is
only beginning beyond depths of $\sim$10.

\begin{figure}
  \includegraphics[bb=60 60 410 800, scale=0.50,angle=0]{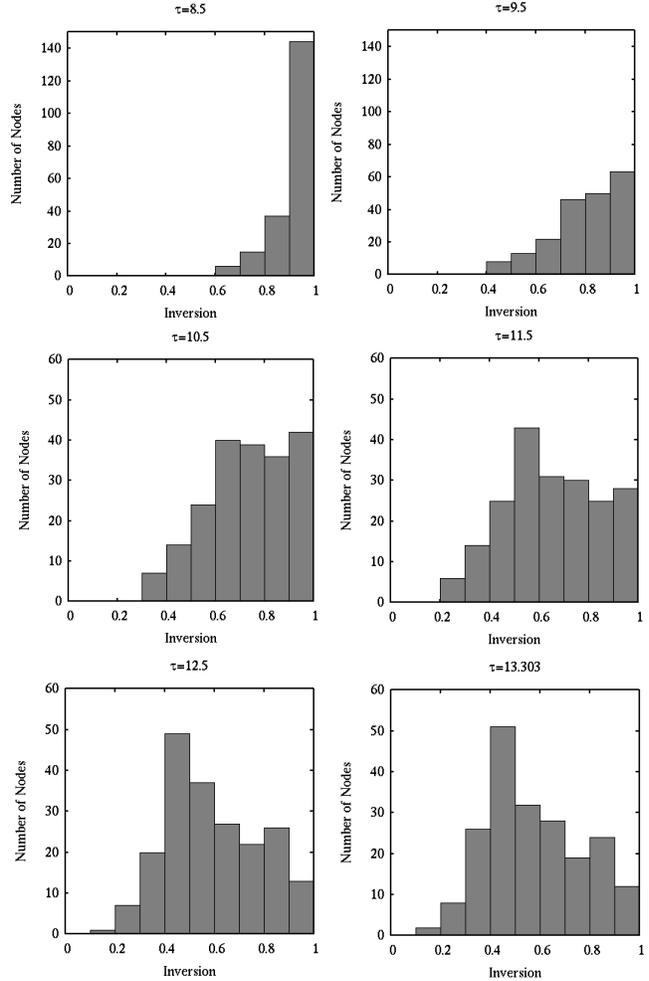}
  \caption{Histograms of the distribution of nodes amongst ten bins representing
inversion ranges. Optical depth multipliers, $\tau$, marked at the top of each
panel increase from left to right along each row, and by rows from top to bottom.
Note the y-axis scale change in moving from the first to the second row.}
\label{f:histogs}
\end{figure}

Figure~\ref{f:histogs} shows the distribution of 
nodal inversions amongst decadal bins with increasing optical depth multiplier. All 202
inversions remained above 0.9 at a depth multiplier of 6.5; the first histogram plotted
as at a depth of 8.5, where the great majority of nodes still have inversions $>$0.9.
With increasing optical depth, an asymmetric distribution is built up, with the most
populated bin moving to progressively lower inversions (rows 2 and 3 in Fig.~\ref{f:histogs}).
The 0.9-1.0 bin ceases to be the most populated between $\tau=10.5$ and $\tau=11.5$, and
at the latter depth, the most populated bin is clearly 0.5-0.6. A further shift to
0.4-0.5 has taken place by $\tau=12.5$.

Concentration of the 
least saturated nodes to the core of the domain, and most saturated nodes
to the boundary, is demonstrated by taking the mean radii of the two groups of
nodes displayed in Fig.~\ref{f:popsat}: the mean radius of the five most saturated
nodes is 0.9856, with a sample standard deviation of 0.0070, whilst the five least saturated nodes 
have a mean radius of only 0.2008 (sample standard deviation of 0.0736).

\subsection{Formal Solutions}
\label{ss:formal}

In the work discussed here, formal solutions were obtained by solving eq.(\ref{eq:rteq})
for a set of rays passing through the domain, with a known gain coefficient, based
on known nodal inversions from eq.(\ref{eq:gcoef}), and see Section~\ref{ss:nodalsol} above.
The nodal solutions can be for any optical depth multiplier, but in most of the work
discussed here, they are for the maximum optical depth multiplier achieved. The
rays for a formal solution originate on a disc that is larger than the maximum extent
of the domain, and the disc was partitioned into zones for each ray as discussed in
Section~\ref{sss:algomain}. Although the nodal solutions are for an undivided inversion, formal
solutions were calculated for a number of spectral bins, or channels. The standard arrangement
in this work allocated 25 spectral bins to cover 7 Doppler widths.

\begin{figure}
  \includegraphics[bb=60 260 410 800, scale=0.48,angle=0]{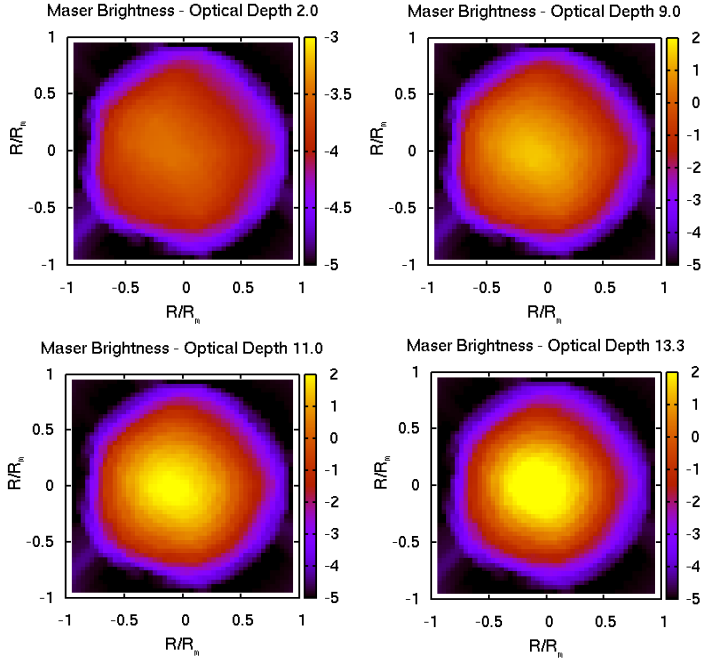}
  \caption{Simulated interferometer images of the cloud viewed from its North pole
           in the brightest (central) spectral channel.
           The value of the optical depth multiplier is shown for each image. On
           the distance scale shown, multiples of the model scale, $R_m$, the observer's
           distance is $10^4$. Note that the figure in the top left, with the optical
           depth multiplier of 2.0, has a different colour scale to the other panels.
           In all cases the colour scale is proportional to the base-10 logarithm of
           the specific intensity divided by the saturation intensity.}
\label{f:formal}
\end{figure}

One of the expected observational features of a spherical, or near-spherical, maser cloud is
that the apparent size of a cloud of fixed dimensions should diminish with increasing
optical depth, a key beaming property discussed, for example, in Chapter~5 of \citet{elitzurbook}.
Testing this property is easy with the code in the current work, since the optical depth
of a cloud of fixed dimensions is determined by a simple multiplier. We show in
Fig.~\ref{f:formal} the same view of the cloud at four different values of the optical
depth multiplier. The observer's position is along the $z$-axis of the model, at a scaled
distance of $10^4$, so the observer is looking down on the North pole of the cloud from
a distance vastly larger than the cloud dimensions, as shown by the $x$ and $y$ scales in
the figure. The image is for the central spectral bin only. 
It is apparent that the emission of the cloud becomes increasingly centre
brightened as the optical depth increases, so that the size of the object, down to, say,
half the maximum brightness, falls as expected. Indeed the effect is more extreme than is
apparent simply from the colour palette in Fig.~\ref{f:formal} because the palette is
re-scaled to a different range of numbers as the brightness range increases. While the
background level is always 10$^{-5}$ with respect to the saturation intensity, the upper
limit changes from 10$^{-3}$ at a depth multiplier of 2.0 to 10$^2$ in the later figures,
where saturation becomes substantial. An observer with an interferometer of a fixed
dynamic range of 10000 would therefore be able to see the object out to a radius of
slightly more than 0.5 in the bottom right-hand panel, but would not detect the object
at the depth of 2.0. The faint plume in the lower left corner of each panel in Fig.~\ref{f:formal}
is an artefact resulting from plotting the circular array of ray origins onto a square
grid.

\begin{figure}
  \includegraphics[bb=60 40 410 300, scale=0.72,angle=0]{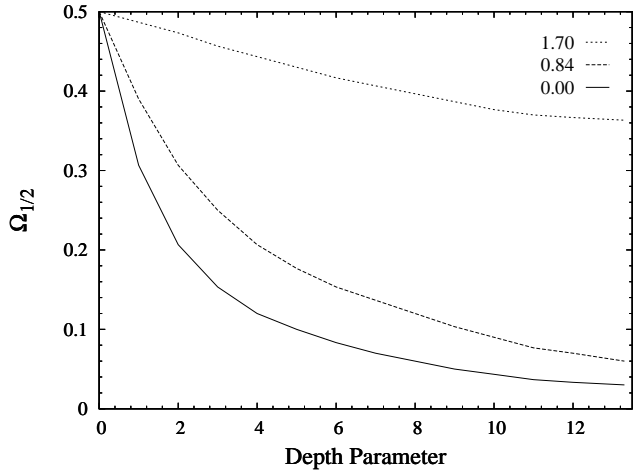}
  \caption{The parameter $\Omega_{1/2}$ as a function of the optical depth multiplier
for three dimensionless frequencies: 0.00, 0.84 and 1.70 Doppler widths from line centre.
The parameter $\Omega_{1/2}$ is the fractional source area or solid angle that contributes half of
the total flux at the observer's position.}
\label{f:area}
\end{figure}

In Fig.~\ref{f:area}, we show the apparent size of the model maser feature as
a function of the optical depth parameter for the line centre and two off-centre
frequencies. The size here is defined as the fractional solid angle, or area, of the source
that produces half the total flux as viewed by the observer, and is represented
as the symbol $\Omega_{1/2}$. It is computed by generating partial fluxes from the
brightest ray, in decreasing order, until half the total flux is reached. The solid
angle corresponding to this subset of rays is then divided by the total solid angle
of the source. The viewpoint and
observer's position are as used for Fig.~\ref{f:formal} above.
In this model, the apparent source size decreases monotonically with increasing maser
depth at all frequencies. The smallest apparent size is found at line centre at all
frequencies and, at any maser depth, the increase in size with frequency offset is
monotonic. Saturation slows the decrease in apparent size with depth at all
frequencies (see Fig.~\ref{f:area}).

\subsection{Simulations of Rotation}
\label{ss:rotn}

In order to study the possibility of flaring events with a single maser cloud, we have
simulated observations of a rotating cloud by viewing a fixed cloud from a number of
regularly spaced positions on a celestial great circle, centered on the cloud.
Corrections were applied via the Doppler effect to convert between
a rotating frame in which the cloud, assumed to be in solid-body rotation, is at rest,
and the inertial frame of the observer. Further consequences of this arrangement with
respect to the input radiation were avoided by assuming a spectrally flat background.
If the rotation period, $P_{dec}$, is measured in decades, and the cloud diameter, $D_{AU}$,
is in Astronomical Units, then the equatorial rotation velocity is
\begin{equation}
v_{rot} = 1.49 D_{AU}/P_{dec} \;\; \mathrm{km\,s^{-1}},
\label{eq:vrot}
\end{equation}
a value comparable to the speed of sound in the cloud gas.

With molecular velocities in the model scaled to a width parameter of
$\bar{W} = \sqrt{2kT/m}$, where $T$ is the kinetic temperature and $m$ is the absolute
molecular mass, rotation of a cloud of diameter $D$ with period $P$ leads to a maximum
Doppler shift, from extreme red to line centre, of
\begin{equation}
\delta_b = 4.14 \frac{D_{AU}}{P_{dec}} \sqrt{\frac{m_u}{T_{100}}}
\label{eq:dopbins}
\end{equation}
frequency bins, where 25 bins cover 7 Doppler widths (see Section~\ref{ss:formal}).
New quantites of order 1 in eq.(\ref{eq:dopbins}) are the molecular mass in atomic
mass units, $m_u$, and the kinetic temperature in units of 100\,K, $T_{100}$. In the
current work, a value of 1 was set for all these parameters, and a total of ten extra
frequency bins assigned to allow for Doppler broadening, Five were allocated either 
side of the line centre, by rounding up the figure of 4.14 bins from eq.(\ref{eq:dopbins}).

The greatest distance between any pair of boundary nodes of the domain was defined to be
the long axis of the cloud. There are obviously an infinite number of unit vectors
perpendicular to the long axis, but a rotation axis was chosen on the basis of the 
minimum absolute component algorithm. Having chosen this particular rotation axis, a set of 
observer's positions were arranged in a plane that passed through the cloud origin
perpendicular to the rotation axis. All observer's positions were placed at a distance
of $10^4$ in scaled units, as for the basic formal solutions. In the current work,
100 observer's positions were used, equally spaced in angle around the plane defined
above. Only one observer's line of sight was exactly parallel to the long axis, and
this occurred at a time of 0.3\,yr.
In the rotation models, the size of the source disc was based on the length of the
long axis regardless of the observer's position, so that the patch of sky viewed subtended
the same solid angle in every case. This is unlike the ordinary formal solutions, where the
size of the source disc is based on the maximum sky-plane extent of the cloud from
the particular observer's position.

Radiation transfer for the formal solutions, one per observer's position, was carried
out in the cloud frame, in which there can be no internal velocity gradients for
solid-body rotation. The
effects of the centrifugal and Coriolis frame forces were ignored, but their effects
are discussed below in Section~\ref{sss:stab}, regarding the stability of the cloud.
Transformation to the observer's frame was effected by calculating the perpendicular sky-plane
distance of each ray from the rotation axis, and computing the associated Doppler
shift in bins. Although the ray paths form a cone with the observer's position at the
apex, the observing positions are sufficiently distant, compared with the
dimensions of the cloud, that convergence of rays through the cloud could be ignored, and
the rays treated as parallel for the purposes of computing Doppler shifts.

The specific intensity of each ray was then shared between a pair of
suitably shifted bins in proportion to the fractional overlap of the single bin
of the old spectrum, and the new pair. The overall effect was mostly rotational
broadening of the spectrum with a small (less than one bin) shift in the peak position
at some observing positions. Overall, the Doppler corrected output radiation makes
observing a fixed cloud from many observers' positions analogous to observing a
rotating cloud from a single position.

\subsubsection{Results and Lightcurves}
\label{sss:lightcurve}

The basic results of the formal solutions simulating rotation are very similar to those of
standard formal solutions: a table of specific intensities and a spectrum of flux densities
for each rotation angle of the cloud. The major difference is that a maser light curve
can be constructed by plotting suitable outputs as a function of rotation angle, or of
time, based on a solid-body rotation speed.

\begin{figure}
  \includegraphics[bb=60 40 410 300, scale=0.72,angle=0]{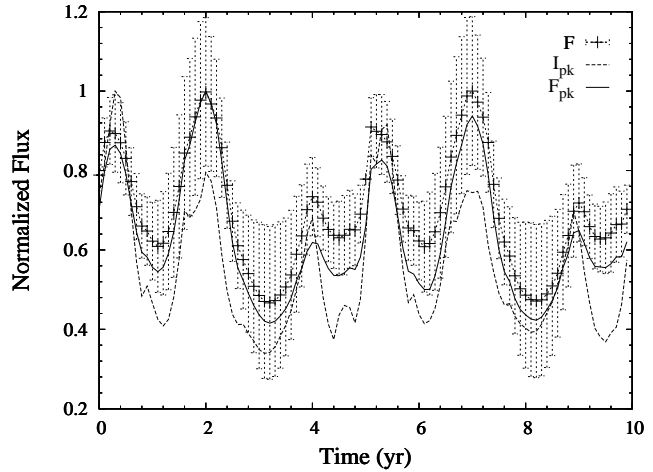}
  \caption{Normalized light curves for three possible observational quantities: $F_{pk}$
(solid line) is the maximum flux density in the spectrum; $I_{pk}$ (dashed line) is the
brightest specific intensity, corresponding to the brightest pixel in an interferometer
map at any channel; $F$ (dotted line) is the flux, integrated over all spectral channels.
All quantities are normalized to the value at their own brightest epoch. Error bars, with
half-height equal to the sample standard deviation, have been added for the integrated
flux curve only for clarity. These standard deviations were computed from 1000 realizations
with rotation axes that were also perpendicular to the long axis of the domain, but were 
otherwise of random orientation.}
\label{f:light}
\end{figure}

There are, perhaps, three useful quantities that can be used on the $y$-axis of a light
curve: one is the brightest specific intensity found in a table of size equal to the
number of rays multiplied by the number of spectral bins. From an observer's point of
view, this would correspond to the brightest pixel in the image cube of a typical
interferometric observation. The second possibility is the maximum flux density in the
spectrum, regardless of the bin in which that maximum occurs. This quantity corresponds
more to results seen by a single-dish observer. The third possibility is the total
(frequency or velocity-integrated) flux. Light curves for all three of these quantities
are plotted in Fig.~\ref{f:light} for a single rotation period of the cloud, 10\,yr.

Perhaps the first point to note from Fig.~\ref{f:light} is that only $I_{pk}$, the brightest
specific intensity, peaks at the time where the long axis is presented directly towards
the observer. The maximum and integrated fluxes instead reach a maximum at 2\,yr, and again,
half a period later, at 7\,yr. All three parameters, however, follow a similar pattern
of four strong maxima per period in two pairs. As expected, the least averaged quantity,
$I_{pk}$, displays the greatest dynamic range: the ratio between its first maximum and
the deepest minimum, at 3.1\,yr, is 2.94. A fainter maximum is visible at $\sim$4\,yr, with
a counterpart at 9\,yr. Rise and fall times for all maxima are of the order of several
months. The pattern of variability is continuous, with no distinct `on' or `off' states.

The curves plotted in Fig.~\ref{f:light} result from rotation about a specific axis that
should not be special except that it was chosen to be perpendicular to the long axis of the
domain. We therefore calculated dispersion information, based on 1000 realizations of the
rotation model, each based on a different rotation axis. Each rotation axis was also
on a circle perpendicular to the long axis, but the angle
around this circle was chosen at random. The chosen dispersion statistic is the sample
standard deviation, and error bars representing this have been plotted, for clarity, on 
the integrated flux curve only. The standard deviations for the other normalized quantities 
have similar behaviour, and are similar in magnitude, but somewhat larger, reflecting the 
less averaged quality of the peak flux and specific intensity. 
At times $t=0$ and $5$\,yr, there are points with very small standard deviations, but
once the observer's position has moved well away from the long axis, typical standard
deviations of 0.15-0.2 were found. This shows that, whilst the peaks near 0 and $5$\,yr
are always present due to orientation of the long axis close to the line of sight, other
peaks and troughs are due only to the particular choice of the rotation axis. It should
be noted, however, that for any given rotation axis, variability of a similar magnitude
to that plotted in Fig.~\ref{f:light} is likely to be observed.

\subsubsection{Stability of the Cloud}
\label{sss:stab}

For any reasonable number density in a maser cloud, the cloud will be gravitationally
stable. The Jeans length is of parsec scale, compared to the AU scale of the cloud. However,
it is still useful to introduce the sound-crossing time, $t_s$, as a non-gravitational
dynamical time, where
\begin{equation}
t_s = D/(2c_s) = 2.2 D_{AU} (m_u/T_{100})^{1/2}\;\;\;{\rm yr},
\label{eq:tcross}
\end{equation}
and the sound speed $c_s = \sqrt{\gamma k T/m}$ has been used with the ratio of specific
heats $\gamma = 7/5$ for rigid diatomic molecules.

If the cloud is uniformly rotating, then, in the rotating frame of the cloud, the dominant forces
acting on a parcel of gas at the surface of the cloud are the centrifugal frame force and
the pressure difference between the cloud and any background medium in which the cloud
is embedded. The equation of motion of the gas for an equatorial parcel of gas may be written,
\begin{equation}
\ddot{r} - \Omega^2 r = -\beta ,
\label{eq:motion}
\end{equation}
where $\beta = \alpha n_X \sigma k T_X / m_X$ is a representation of the acceleration imparted
by the pressure gradient, which has been written as $\alpha$ multiplied by the pressure of
the external medium of number density, temperature and molecular mass, respectively $n_X$,
$T_X$, $m_X$, acting over the mean-free-path $1/(\sigma n_X)$, where sigma is the collision
cross section. The general solution to eq.(\ref{eq:motion}), if the initial radial velocity
of the parcel of gas is zero, and the initial radius is the cloud radius, $R$, is
\begin{equation}
r(t) = (R - \beta / \Omega^2) \cosh (\Omega t) +  \beta / \Omega^2 .
\label{eq:soln}
\end{equation}
Two interesting extremes of eq.(\ref{eq:soln}) are firstly the case of negligible external pressure
($\beta \sim 0$) in which case the cloud is unstable and disperses on a timescale of order 
$1/\Omega = P/(2\pi)$: for the cloud modelled in Fig.~\ref{f:light}, approximately 1.6\,yr.
Secondly, in the case where $R-\beta/\Omega^2 \ll 1$, the first term in eq.(\ref{eq:soln}) can
be ignored until some much larger time, and for times up to at least the order of $P$, the
radius remains close to the constant value of $\beta / \Omega^2 \sim R$. In the second
case, the cloud is stable in the short term provided that $\alpha$ is small, that is the
inward acceleration for centrifugal balance can be provided by only a small fraction of the 
pressure of the gas external to the cloud. With $\beta = R\Omega^2$, the value of $\alpha$ is
\begin{equation}
\alpha = \left( \frac{V_{rot}}{c_{sX}} \right)^2 \frac{1}{R\sigma n_X}
       \simeq 7.3\times 10^{-4} \frac{D_{AU}}{P_{dec}^2 T_{1000} \sigma_{20} n_6},
\label{eq:alpha}
\end{equation}
where $V_{rot} = \Omega R$ and $c_{sX}$ is the sound speed in the external gas. Other
parameters are $T_{1000}$, the external gas temperature in units of 1000\,K, $n_6$, the
number density of the same material in units of $10^6$\,cm$^{-3}$, and $\sigma_{20}$, the
collision cross-section in units of $10^{-20}$\,m$^{2}$. It does appear from eq.(\ref{eq:alpha})
that $\alpha$ will usually be small, allowing for a number of rotations before the cloud
is destroyed.

Observational evidence from long-term single-dish monitoring of H$_2$O masers
in massive star-forming regions 
\citep{1999ARep...43..209L}, suggests that, in a turbulence interpretation, there
are either no long-lived vortices (which survive many rotations), or that the
power spectrum of the turbulence is concentrated towards the largest scales. This
in turn
suggests that significant evolution in the shape of a cloud is likely in a single
revolution, although identification of the model cloud with a turbulent eddy may
not generally be appropriate. In evolved-stars, however, some maser structures supporting
the 1612-MHz OH line survive for several pulsational cycles of
typically 1\,yr \citep{2000A&AS..146..179E}, and may therefore be long-lived if
they are turbulence based.

\subsection{Execution Times}
\label{timings}

Calculation of the nodal solutions dominates the execution time and,
for the version of the code used in the present work, this is in turn dominated by the need
to compute a Jacobian matrix of the derivatives of eq.(\ref{eq:globdisc}) for the LM
algorithm (see Section~\ref{sss:algomain}). The time needed to compute the Jacobian 
scales as the square of the number of nodes, whilst other calls to the solution checker 
scale only linearly. The time to complete one iteration of the Levenberg-Marquardt 
algorithm for the domain discussed in this work, call it
$T$, is therefore approximately the time to compute the Jacobian, and it is equal to
$\sim$270\,s on an Intel Core i7-3930K CPU, rated at 3.20\,GHz.

For values of the optical depth multiplier $\tau <$10, each solution requires an
average of 2-3$T$, rising to $\sim$12$T$ when $\tau$ is in
the approximate range 10-11.5. These estimates are based on a step of 0.1 in $\tau$
between solutions. This value of the step was used to give smooth results in
graphs such as Fig.~\ref{f:popsat}, and larger steps could have been used at small $\tau$.
For values of $\tau >$11.5, progressively smaller step sizes must be used, 
and these became prohibitively small
near the largest value of $\tau$ used in current work (13.303).

\section{Discussion}
\label{discuss}

There have been several previous essays at 3D-modelling of astrophysical masers, for
example the toroidal model of hydrogen radio recombination-line masers in MWC349A
\citep{2013A&A...553A..45B}, and the protostellar (T-Tauri) disc model of methanol masers
\citep{2016MNRAS.460.2648P}. However,
the results presented in this paper are probably the first fully-3D maser solutions for an arbitrary
degree of saturation. They span a range of stages from computation of
inversions at each node of a triangulated domain, through the use of those solutions to
carry out formal radiative transfer of rays from a background source to an observer, and
the use of modified formal solutions to generate light-curves for a uniformly rotating
maser cloud with Doppler rotational broadening.

Nodal solutions follow the expected pattern for an approximately spherical cloud, with
the most saturated nodes found near the edges of the domain, and the least saturated
nodes located near the centre. The brightest maser rays have intensities of approximately
600 times the saturation intensity in the most highly saturated version of the model.
Formal solutions through a domain with known saturated nodal inversions leads to a clear
reduction in the observed source size with increasing amplification and saturation.
The effective source size, quantified as the source solid angle that provides half the
observed flux, is clearly a function of frequency, and decreases monotonically with
the degree of saturation at all frequencies. No effects such as the decrease in
effective size with frequency for detunings smaller than approximately 1 Doppler
width, as in \citet{1994ApJ...424..991E} were observed, but this is expected since
the current work uses a CVR, rather than NVR, approximation for redistribution.

The light-curves show that very small deviations from spherical symmetry in a cloud
supporting intense maser emission will result in significant changes in observed brightnesses,
flux densities and fluxes. However, for cloud of similar shape to the one
used here, the radiation beaming
pattern is still far from the ideal filamentary type, even for the maximum depth parameter
used. The contrast between minimum and maximum peak brightness for the rotation simulation is a 
factor of $\sim$3. Whilst this behaviour might be described as a 
small-amplitude flaring event, it can certainly not match the dynamic ranges of hundreds to
thousands seen in, for example, extreme water-maser flares. Although the light-curves plotted
in Fig.~\ref{f:light} are periodic, we would expect that geometrical evolution of the
cloud would make variability from a real rotating cloud system only quasi-periodic. 
The dynamic ranges from the model light curve are more in accord with many of the
flaring methanol sources monitored by \citet{2004MNRAS.355..553G} over $\sim$4\,yr.
Two examples for comparison might be G331.13-024 (periodic) and G351.78-0.54 (aperiodic).

The numerical procedure used in this work has demonstrated that the integral equation
method of solving the combined radiative transfer and inversion saturation problem
is viable in 3-D finite element domains for strongly saturated maser sources. The method 
may be readily generalised to multi-level systems, weaker redistribution and non-uniform clouds.
In the latter case, a more conventional numerical frequency integration is almost
certainly better than the analytical technique used here, though under NVR this issue
does not arise. A $J=1-0$ Zeeman polarization version of this model is under construction
(Etoka \& Gray, in preparation). For the more general multi-level case, we will work
with fractional energy-level populations as the unknowns, and solve the set of
statistical master equations for these populations. The coefficients of the master
equations contain mean intensities that can be eliminated, as in the present work.
However, in the general case, the mean intensity of a transition is replaced by a
depth, solid angle and frequency integral over the pair of populations involved
in the transition.

As noted in Section~\ref{timings}, performance of the code significantly degrades at
values of the optical depth multiplier $>$11.5. This does not appear to be due to oscillation
between multiple solutions of the non-linear equation problem, but instead appears
to be due to an increasingly `shallow' global minimum in the residuals. The fact that our
strategy of starting from previous solutions at lower optical depth (see Section~\ref{sss:algomain})
is effective suggests that this is indeed the case. Improving the
convergence properties of the code at high optical depths is obviously highly desirable. We
have evidence that, as expected, increased resolution (more nodes) will do this, but
at the expense of increased computing time. Other techniques that we intend to consider
are better optimisation of the aspect ratios of the finite elements, better extrapolation
algorithms than the current {\sc polint}, and linear
perturbation of eq.(\ref{eq:globdisc}). 

\section{Conclusions}
\label{conclusion}
(1) Nodal inversion and formal RT solutions are as expected for a near-spherical
maser; a discussion of historical apparent size versus frequency effects requires
extension to an NVR model.
(2) Rotation of the cloud provides a periodic signal of dynamic range approximately
3, purely from geometry, even for a domain that has only small departures from
spherical symmetry.
(3) On the grounds that a real cloud would likely evolve significantly during
one rotation, this model cannot explain periodic flaring, but could produce
quasi-periodic variability.
(4) The domain used in the current work yields a light-curve that is not strongly
weighted to either the higher or to the lower part of the flux range. It is
therefore not a good model of sources with a very high, or very low, duty cycle.

\section*{Acknowledgments}

MDG and SE acknowledge funding from the UK Science and Technology Facilities
Council (STFC) as part of the consolidated grant ST/P000649/1 to the Jodrell Bank
Centre for Astrophysics at the University of Manchester. LM would like to
thank the Nuffield Foundation for a summer bursary. We would like to thank the
anonymous referee for helpful comments.

\bibliographystyle{mn2e}
\bibliography{MDGray_MN17}

\begin{thebibliography}{}

\bibitem[\protect\citeauthoryear{{Abraham}, {Opher} \& {Raffaelli}}{{Abraham}
  et~al.}{1979}]{1979IAUC.3415....2A}
{Abraham} Z.,  {Opher} R.,    {Raffaelli} J.~C.,  1979, \iaucirc, 3415

\bibitem[\protect\citeauthoryear{{Abramowitz} \& {Stegun}}{{Abramowitz} \&
  {Stegun}}{1972}]{absteg}
{Abramowitz} M.,  {Stegun} I.~A.,  1972, {Handbook of Mathematical Functions}

\bibitem[\protect\citeauthoryear{{Alcock} \& {Ross}}{{Alcock} \&
  {Ross}}{1985}]{1985ApJ...290..433A}
{Alcock} C.,  {Ross} R.~R.,  1985, \apj, 290, 433

\bibitem[\protect\citeauthoryear{Amini \& Rostami}{Amini \&
  Rostami}{2015}]{Amini2015341}
Amini K.,  Rostami F.,  2015, Journal of Computational and Applied Mathematics,
  288, 341

\bibitem[\protect\citeauthoryear{{Anderson} \& {Watson}}{{Anderson} \&
  {Watson}}{1993}]{1993ApJ...407..620A}
{Anderson} N.,  {Watson} W.~D.,  1993, \apj, 407, 620

\bibitem[\protect\citeauthoryear{{Araya}, {Hofner}, {Sewi{\l}o}, {Linz},
  {Kurtz}, {Olmi}, {Watson} \& {Churchwell}}{{Araya}
  et~al.}{2007}]{2007ApJ...654L..95A}
{Araya} E.,  {Hofner} P.,  {Sewi{\l}o} M.,  {Linz} H.,  {Kurtz} S.,  {Olmi} L.,
   {Watson} C.,    {Churchwell} E.,  2007, \apjl, 654, L95

\bibitem[\protect\citeauthoryear{{Argon}, {Greenhill}, {Reid}, {Moran},
  {Menten}, {Henkel} \& {Inoue}}{{Argon} et~al.}{1993}]{1993AAS...182.7507A}
{Argon} A.~L.,  {Greenhill} L.~J.,  {Reid} M.~J.,  {Moran} J.~M.,  {Menten}
  K.~M.,  {Henkel} C.,    {Inoue} M.,  1993, in American Astronomical Society
  Meeting Abstracts \#182 Vol.~25 of Bulletin of the American Astronomical
  Society, {First VLBI Observations of Extragalactic Water Vapor Maser Flares
  in IC10}.
p.~926

\bibitem[\protect\citeauthoryear{{B{\'a}ez-Rubio}, {Mart{\'{\i}}n-Pintado},
  {Thum} \& {Planesas}}{{B{\'a}ez-Rubio} et~al.}{2013}]{2013A&A...553A..45B}
{B{\'a}ez-Rubio} A.,  {Mart{\'{\i}}n-Pintado} J.,  {Thum} C.,    {Planesas} P.,
   2013, \aap, 553, A45

\bibitem[\protect\citeauthoryear{Beckers \& Beckers}{Beckers \&
  Beckers}{2012}]{Beckers2012275}
Beckers B.,  Beckers P.,  2012, Computational Geometry, 45, 275

\bibitem[\protect\citeauthoryear{{Bettwieser}}{{Bettwieser}}{1976}]{1976A&A...%
.50..231B}
{Bettwieser} E.,  1976, \aap, 50, 231

\bibitem[\protect\citeauthoryear{{Boboltz}, {Simonetti}, {Dennison}, {Diamond}
  \& {Uphoff}}{{Boboltz} et~al.}{1998}]{1998ApJ...509..256B}
{Boboltz} D.~A.,  {Simonetti} J.~H.,  {Dennison} B.,  {Diamond} P.~J.,
  {Uphoff} J.~A.,  1998, \apj, 509, 256

\bibitem[\protect\citeauthoryear{{Brinch} \& {Hogerheijde}}{{Brinch} \&
  {Hogerheijde}}{2010}]{2010A&A...523A..25B}
{Brinch} C.,  {Hogerheijde} M.~R.,  2010, \aap, 523, A25

\bibitem[\protect\citeauthoryear{{Cimerman}}{{Cimerman}}{1979}]{1979ApJ...228L%
..79C}
{Cimerman} M.,  1979, \apjl, 228, L79

\bibitem[\protect\citeauthoryear{{Daniel} \& {Cernicharo}}{{Daniel} \&
  {Cernicharo}}{2013}]{2013A&A...553A..70D}
{Daniel} F.,  {Cernicharo} J.,  2013, \aap, 553, A70

\bibitem[\protect\citeauthoryear{{Elitzur}}{{Elitzur}}{1990}]{1990ApJ...363..6%
38E}
{Elitzur} M.,  1990, \apj, 363, 638

\bibitem[\protect\citeauthoryear{{Elitzur}}{{Elitzur}}{1992}]{elitzurbook}
{Elitzur} M.,  1992, {Astronomical masers}.
Kluwer, Dordrecht

\bibitem[\protect\citeauthoryear{{Elitzur} \& {Asensio Ramos}}{{Elitzur} \&
  {Asensio Ramos}}{2006}]{2006MNRAS.365..779E}
{Elitzur} M.,  {Asensio Ramos} A.,  2006, \mnras, 365, 779

\bibitem[\protect\citeauthoryear{{Emmering} \& {Watson}}{{Emmering} \&
  {Watson}}{1994}]{1994ApJ...424..991E}
{Emmering} R.~T.,  {Watson} W.~D.,  1994, \apj, 424, 991

\bibitem[\protect\citeauthoryear{{Etoka} \& {Le Squeren}}{{Etoka} \& {Le
  Squeren}}{1996}]{1996A&A...315..134E}
{Etoka} S.,  {Le Squeren} A.~M.,  1996, \aap, 315, 134

\bibitem[\protect\citeauthoryear{{Etoka} \& {Le Squeren}}{{Etoka} \& {Le
  Squeren}}{2000}]{2000A&AS..146..179E}
{Etoka} S.,  {Le Squeren} A.~M.,  2000, \aaps, 146, 179

\bibitem[\protect\citeauthoryear{{Goedhart}, {Gaylard} \& {van der
  Walt}}{{Goedhart} et~al.}{2003}]{2003MNRAS.339L..33G}
{Goedhart} S.,  {Gaylard} M.~J.,    {van der Walt} D.~J.,  2003, \mnras, 339,
  L33

\bibitem[\protect\citeauthoryear{{Goedhart}, {Gaylard} \& {van der
  Walt}}{{Goedhart} et~al.}{2004}]{2004MNRAS.355..553G}
{Goedhart} S.,  {Gaylard} M.~J.,    {van der Walt} D.~J.,  2004, \mnras, 355,
  553

\bibitem[\protect\citeauthoryear{{Goedhart}, {Minier}, {Gaylard} \& {van der
  Walt}}{{Goedhart} et~al.}{2005}]{2005MNRAS.356..839G}
{Goedhart} S.,  {Minier} V.,  {Gaylard} M.~J.,    {van der Walt} D.~J.,  2005,
  \mnras, 356, 839

\bibitem[\protect\citeauthoryear{{Goldreich} \& {Keeley}}{{Goldreich} \&
  {Keeley}}{1972}]{1972ApJ...174..517G}
{Goldreich} P.,  {Keeley} D.~A.,  1972, \apj, 174, 517

\bibitem[\protect\citeauthoryear{{Gray}}{{Gray}}{2012}]{mybook}
{Gray} M.~D.,  2012, {Maser Sources in Astrophysics}.
Cambridge University Press, Cambridge, UK

\bibitem[\protect\citeauthoryear{{King} \& {Florance}}{{King} \&
  {Florance}}{1964}]{1964ApJ...139..397K}
{King} J.~I.~F.,  {Florance} E.~T.,  1964, \apj, 139, 397

\bibitem[\protect\citeauthoryear{{Lang} \& {Bender}}{{Lang} \&
  {Bender}}{1973}]{1973ApJ...180..647L}
{Lang} R.,  {Bender} P.~L.,  1973, \apj, 180, 647

\bibitem[\protect\citeauthoryear{{Lekht}, {Mendoza-Torres} \&
  {Silant'ev}}{{Lekht} et~al.}{1999}]{1999ARep...43..209L}
{Lekht} E.~E.,  {Mendoza-Torres} J.~E.,    {Silant'ev} N.~A.,  1999, Astronomy
  Reports, 43, 209

\bibitem[\protect\citeauthoryear{{Lekht} \& {Sorochenko}}{{Lekht} \&
  {Sorochenko}}{1984}]{1984SvAL...10..307L}
{Lekht} E.~E.,  {Sorochenko} R.~L.,  1984, Soviet Astronomy Letters, 10, 307

\bibitem[\protect\citeauthoryear{{Litvak}}{{Litvak}}{1973}]{1973ApJ...182..711%
L}
{Litvak} M.~M.,  1973, \apj, 182, 711

\bibitem[\protect\citeauthoryear{{Madden}, {Friberg}, {Brown} \&
  {Godfrey}}{{Madden} et~al.}{1988}]{1988IAUC.4537....1M}
{Madden} S.~C.,  {Friberg} P.,  {Brown} R.,    {Godfrey} P.,  1988, \iaucirc,
  4537, 1

\bibitem[\protect\citeauthoryear{{Neufeld}}{{Neufeld}}{1992}]{1992ApJ...393L..%
37N}
{Neufeld} D.~A.,  1992, \apjl, 393, L37

\bibitem[\protect\citeauthoryear{{Ng}}{{Ng}}{1974}]{1974JChPh..61.2680N}
{Ng} K.-C.,  1974, \jcp, 61, 2680

\bibitem[\protect\citeauthoryear{{Olson} \& {Kunasz}}{{Olson} \&
  {Kunasz}}{1987}]{1987JQSRT..38..325O}
{Olson} G.~L.,  {Kunasz} P.~B.,  1987, \jqsrt, 38, 325

\bibitem[\protect\citeauthoryear{{Parfenov}, {Semenov}, {Sobolev} \&
  {Gray}}{{Parfenov} et~al.}{2016}]{2016MNRAS.460.2648P}
{Parfenov} S.~Y.,  {Semenov} D.~A.,  {Sobolev} A.~M.,    {Gray} M.~D.,  2016,
  \mnras, 460, 2648

\bibitem[\protect\citeauthoryear{{Press}, {Teukolsky}, {Vetterling} \&
  {Flannery}}{{Press} et~al.}{1992}]{1992nrfa.book.....P}
{Press} W.~H.,  {Teukolsky} S.~A.,  {Vetterling} W.~T.,    {Flannery} B.~P.,
  1992, {Numerical recipes in FORTRAN. The art of scientific computing}

\bibitem[\protect\citeauthoryear{{Rajabi} \& {Houde}}{{Rajabi} \&
  {Houde}}{2017}]{2017SciA....3E1858R}
{Rajabi} F.,  {Houde} M.,  2017, Science Advances, 3, e1601858

\bibitem[\protect\citeauthoryear{{Scappaticci} \& {Watson}}{{Scappaticci} \&
  {Watson}}{1992}]{1992ApJ...387L..73S}
{Scappaticci} G.~A.,  {Watson} W.~D.,  1992, \apjl, 387, L73

\bibitem[\protect\citeauthoryear{{Spaans} \& {van Langevelde}}{{Spaans} \& {van
  Langevelde}}{1992}]{1992MNRAS.258..159S}
{Spaans} M.,  {van Langevelde} H.~J.,  1992, \mnras, 258, 159

\bibitem[\protect\citeauthoryear{{Strelnitskii}}{{Strelnitskii}}{1982}]{1982Sv%
AL....8...86S}
{Strelnitskii} V.~S.,  1982, Soviet Astronomy Letters, 8, 86

\bibitem[\protect\citeauthoryear{{Sugiyama}, {Nagase}, {Yonekura}, {Momose},
  {Yasui}, {Saito}, {Motogi}, {Honma}, {Hachisuka}, {Matsumoto}, {Uchiyama} \&
  {Fujisawa}}{{Sugiyama} et~al.}{2017}]{2017PASJ...69...59S}
{Sugiyama} K.,  {Nagase} K.,  {Yonekura} Y.,  {Momose} M.,  {Yasui} Y.,
  {Saito} Y.,  {Motogi} K.,  {Honma} M.,  {Hachisuka} K.,  {Matsumoto} N.,
  {Uchiyama} M.,    {Fujisawa} K.,  2017, \pasj, 69, 59

\bibitem[\protect\citeauthoryear{{Sullivan}
  III}{{Sullivan}}{1973}]{1973ApJS...25..393S}
{Sullivan} III W.~T.,  1973, \apjs, 25, 393

\bibitem[\protect\citeauthoryear{{Szymczak}, {Olech}, {Wolak}, {Bartkiewicz} \&
  {Gawro{\'n}ski}}{{Szymczak} et~al.}{2016}]{2016MNRAS.459L..56S}
{Szymczak} M.,  {Olech} M.,  {Wolak} P.,  {Bartkiewicz} A.,    {Gawro{\'n}ski}
  M.,  2016, \mnras, 459, L56

\bibitem[\protect\citeauthoryear{{Watson} \& {Wyld}}{{Watson} \&
  {Wyld}}{2003}]{2003ApJ...598..357W}
{Watson} W.~D.,  {Wyld} H.~W.,  2003, \apj, 598, 357

\bibitem[\protect\citeauthoryear{Zienkiewicz \& Taylor}{Zienkiewicz \&
  Taylor}{2000}]{zienkiebook}
Zienkiewicz O.,  Taylor R.,  2000, The Finite Element Method: The basis.
Butterworth-Heinemann

\end{thebibliography}

\appendix

\section[]{Analytic Frequency Integration}
\label{app:freq}

The standard power series expansion of the exponential in braces on the second 
line of eq.(\ref{eq:nlint}) leads to a general term, for power $n$ of the form
\begin{equation}
q_n = (1/n!)\left[
                    \int_{\vec{r}_0}^{\vec{r}}
                    d\vec{r}' \Delta(\vec{r}') \eta (\vec{r}')
                    e^{-\eta^2(\vec{r}')(\tilde{\nu} - \hat{\vec{n}} \cdot \vec{u}(\vec{r}'))^2}
            \right]^n.
\label{eq:appA1}
\end{equation}
All such terms are subject to the frequency and solid angle integrals that
appear in eq.(\ref{eq:nlint}). The power form of the expression in eq.(\ref{eq:appA1})
may be expanded to the multiple integral form,
\begin{align}
q_n &= (1/n!)\int_{\vec{r}_0}^{\vec{r}} d\vec{r}_n \Delta(\vec{r}_n) \eta(\vec{r}_n)
             \int_{\vec{r}_0}^{\vec{r}} d\vec{r}_{n-1} \Delta(\vec{r}_{n-1}) \eta(\vec{r}_{n-1})... \nonumber \\
           &...  \int_{\vec{r}_0}^{\vec{r}} d\vec{r}_1 \Delta(\vec{r}_1) \eta(\vec{r}_1)
               \exp \left[
                   -\sum_{k=1}^n \eta^2(\vec{r}_k)(\tilde{\nu} - \hat{\vec{n}}\cdot \vec{u}(\vec{r}_k))^2
                    \right],
\label{eq:appA2}
\end{align}
noting that only the final exponential term is a function of frequency. The frequency
integral from eq.(\ref{eq:nlint}) can therefore be brought inside all the spatial
integrals in eq.(\ref{eq:appA2}) and written as,
\begin{equation}
{\cal F} = \int_{-\infty}^\infty d\tilde{\nu} \exp \left[
                   -\sum_{k=0}^n \eta^2(\vec{r}_k)(\tilde{\nu} - \hat{\vec{n}}\cdot \vec{u}(\vec{r}_k))^2
                                     \right],
\label{eq:appA3}
\end{equation}
where $\vec{r}_0 = \vec{r}$, the final position of the ray from eq.(\ref{eq:nlint}).
The integral in eq.(\ref{eq:appA3}) can be carried out analytically in general via 
formula 7.4.32 from
\citet{absteg} with the coefficients $a=\sum_{k=0}^n \eta^2(\vec{r}_k)$ of the squared frequency, 
$b= -\sum_{k=0}^n \hat{\vec{n}}\cdot \vec{u}(\vec{r}_k) \eta^2(\vec{r}_k)$ of
the frequency, and
the frequency-independent term $c= \sum_{k=0}^n ( \hat{\vec{n}}\cdot \vec{u}(\vec{r}_k))^2 \eta^2(\vec{r}_k)$. 
The result cannot generally be re-factored,
but if there is no velocity field inside the cloud ($\vec{u}=(0,0,0)$ everywhere),
then $b=c=0$ and the result of eq.(\ref{eq:appA3}) is simply,
\begin{equation}
{\cal F} = \left(
              \frac{\pi}{\sum_{k=0}^n \eta^2(\vec{r}_k)} 
           \right)^{1/2} .
\label{eq:appA4}
\end{equation}
In the case of a uniform cloud, which will be assumed from now on, $\eta = 1$ everywhere,
and eq.(\ref{eq:appA4}) therefore reduces
to ${\cal F} = \pi^{1/2} / \sqrt(n+1)$. This expression is a constant that can be
removed from all the spatial integrals in eq.(\ref{eq:appA2}), leaving them in a
form that is already factored, and can be restored to the power form in eq.(\ref{eq:termn}) of
the main text.

\end{document}